\documentclass[aps,prl,twocolumn,superscriptaddress]{revtex4}
\usepackage{mathrsfs}
\usepackage{epsfig}
\usepackage{graphicx}
\usepackage{amsfonts}
\usepackage[figuresright]{rotating}
\usepackage{amssymb}
\usepackage{amsmath}
\usepackage{dcolumn}
\usepackage{bm}
\usepackage{color}

\def\Im{\rm{Im}}
\def\be{\begin{equation}} \def\ee{\end{equation}}
\def\bea{\begin{eqnarray}} \def\eea{\end{eqnarray}}

\newcommand{\CIT} {Department of Physics and Jiangsu Laboratory of Advanced Functional Material, Changshu Institute of Technology, Changshu 215500, China}

\newcommand{\sjtu} {Key Laboratory of Artificial Structures and Quantum
Control, Department of Physics and Astronomy, Shanghai Jiao Tong University, Shanghai 200240, People's Republic of China}

\newcommand{\CSRC} {Beijing Computational Science Research Center, Beijing 100193, China}

\newcommand{\WQCASQC} {Wilczek Quantum Center, School of Physics and Astronomy, Shanghai Jiao Tong University, Shanghai 200240, China}

\newcommand{\tdli}{Shenyang National Laboratory for Materials Science and T. D. Lee Institute, Shanghai Jiao Tong University, Shanghai 200240, China}

\newcommand{\BHU}{
School of Physics, Key Laboratory of Micro-Nano Measurement-Manipulation
and Physics (Ministry of Education), Beihang University, Beijing 100191, China }

\begin{document}
\title{Noise-driven universal dynamics towards an infinite temperature state}

\author{Jie Ren}
\affiliation{\CIT}
\affiliation{\sjtu}

\author{Qiaoyi Li}
\affiliation{\BHU}

\author{Wei Li}
\affiliation{\BHU}

\author{Zi Cai}
\email{zcai@sjtu.edu.cn}
\affiliation{\WQCASQC}
\affiliation{\sjtu}

\author{Xiaoqun Wang}
\affiliation{\sjtu}
\affiliation{\CSRC}
\affiliation{\tdli}

\begin{abstract}  
Dynamical universality is the observation that the dynamical properties of different systems might exhibit universal behavior that are independent of the system details. In this paper, we study the long-time dynamics of an one-dimensional noisy quantum magnetic model, and find that even though the system are inevitably driven to an infinite temperature state, the relaxation dynamics towards such featureless state can be highly nontrivial and universal.  The effect of various mode-coupling mechanisms (external potential, disorder, interaction, and the interplay between them) as well as the conservation law on the long-time dynamics of the systems have been studied, and their relevance with current ultracold atomic experiments have been discussed.
\end{abstract}


\maketitle

{\it Introduction:} One of the most striking phenomena in physics is that systems with radically diverse microstructures can exhibit universal properties independent of system details. For instance, near the critical points, various systems might share similar relaxation behavior (e.g. the same dynamical critical exponents $z$\cite{Hohenberg1977}). Such dynamical universality, in general, is more complex and richer than the static one, because it is not only determined by  conventional factors such as symmetries, dimensionality, etc., but is also affected by other factors (e.g.conservation laws)  that are irrelevant for static criticality\cite{Taeuber2014}. Furthermore, they can also be observed in systems far from equilibrium (the Kardar-Parisi-Zhang(KPZ) model\cite{Kardar1986} for instance) other than in the vicinity of equilibrium critical points. So far, the majority of studies are restricted to classical systems, while the dynamical universality in quantum systems is far from being understood due to the intrinsic difficulty of dealing with non-equilibrium quantum many-body systems.

In this study, we investigate the universal relaxation dynamics of quantum many-body systems subjected to white noises. In the absence of energy dissipation, the system will be continuously heated by noises and eventually driven into an infinite temperature state(ITS). Despite the triviality of such steady state, the way to approach it can be highly nontrivial\cite{Marino2012,Poletti2012,Poletti2013,Pichler2013,Cai2013,Sieberer2013,Buchhold2015}. For closed quantum many-body systems without noises, the dynamics near the ITS  can be either diffusive or super-diffusive depending on the integrability of the systems\cite{Ljubotina2019,Nardis2019,Dupont2019},  while for the open quantum systems as we will show, the noise-driven relaxation dynamics is more complex and richer.  It can be either algebraic, exponential or stretched exponential decay depending on the parameters of the systems. Even for the same parameters, different physical quantities might exhibit distinct relaxation dynamics. It is possible that some of them are purely diffusive while at the same time, the relaxation dynamics of other quantities are neither diffusive nor super-diffusive.

The model we studied is an one-dimensional(1D) noisy quantum magnetic model with disorder. A model without noises has been intensively studied in the context of many-body localization(MBL)\cite{Basko2006,Oganesyan2007,Znidaric2008,Pal2010,Schreiber2015}. Even with noises,  similar models have been studied to explore the stability of MBL against decoherence, which is only relevant with the short time dynamics in a strong disorder regime\cite{Levi2016,Fischer2016,Medvedyeva2016,Everest2017,Wu2019}. However, as we will show, to observe the universal dynamics of interest, one needs a system large enough to eliminate the strong finite-size effect, waits for sufficiently long time until the initial state information is washed out, and focuses on the weak disorder regime, which exhibits interesting transport behavior even in the absence of noises\cite{Znidaric2016}. Here, we systematically studied the relaxation dynamics of two different kinds of physical quantities, and focus on the dynamical universality and their dependence on various mode-coupling mechanisms (disorder, interaction and so on) and the conservation laws.

\begin{figure*}[htb]
\includegraphics[width=0.32\linewidth,bb=76 49 714 510]{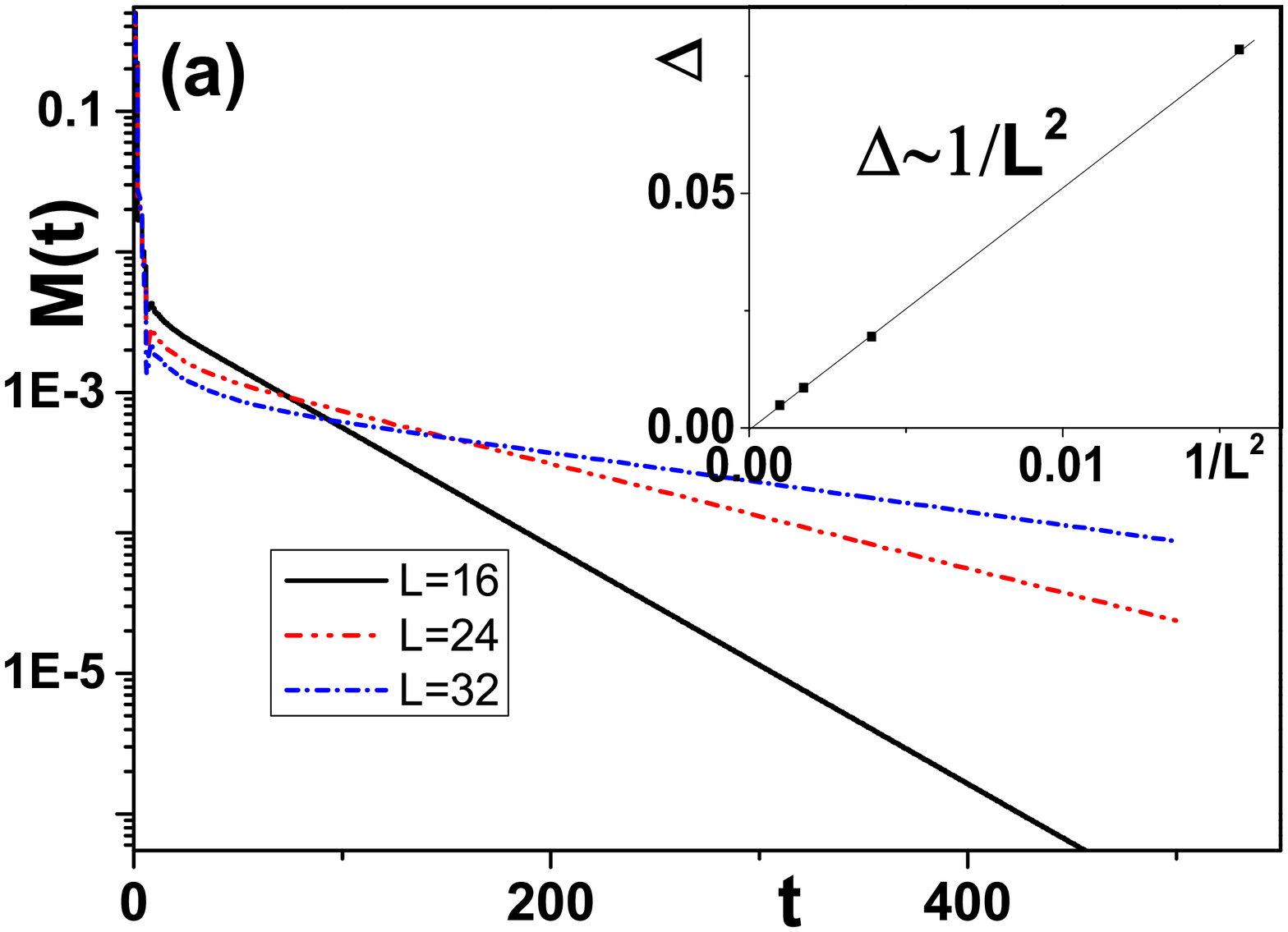}
\includegraphics[width=0.32\linewidth,bb=80 59 732 531]{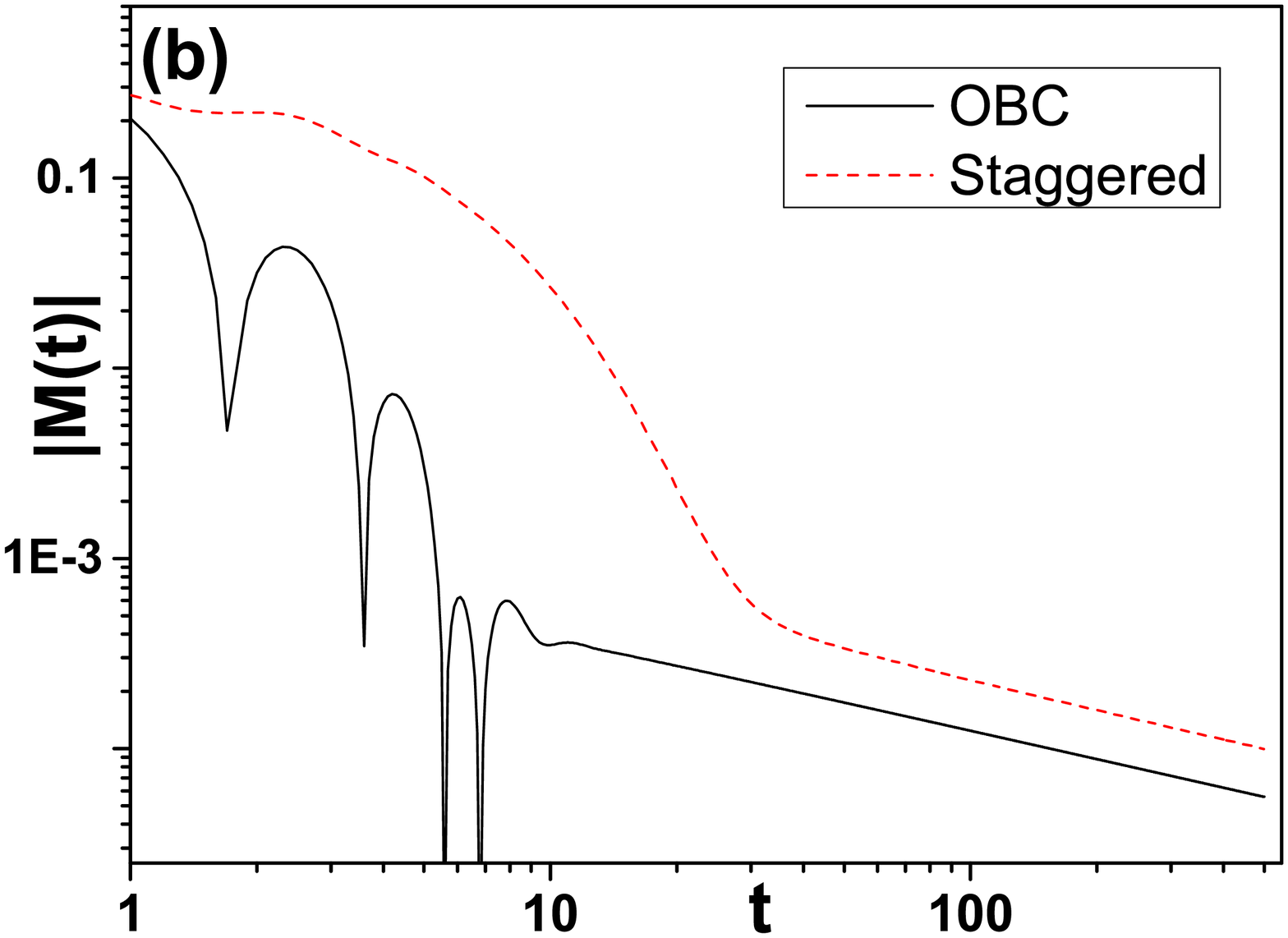}
\includegraphics[width=0.33\linewidth,bb=90 58 733 531]{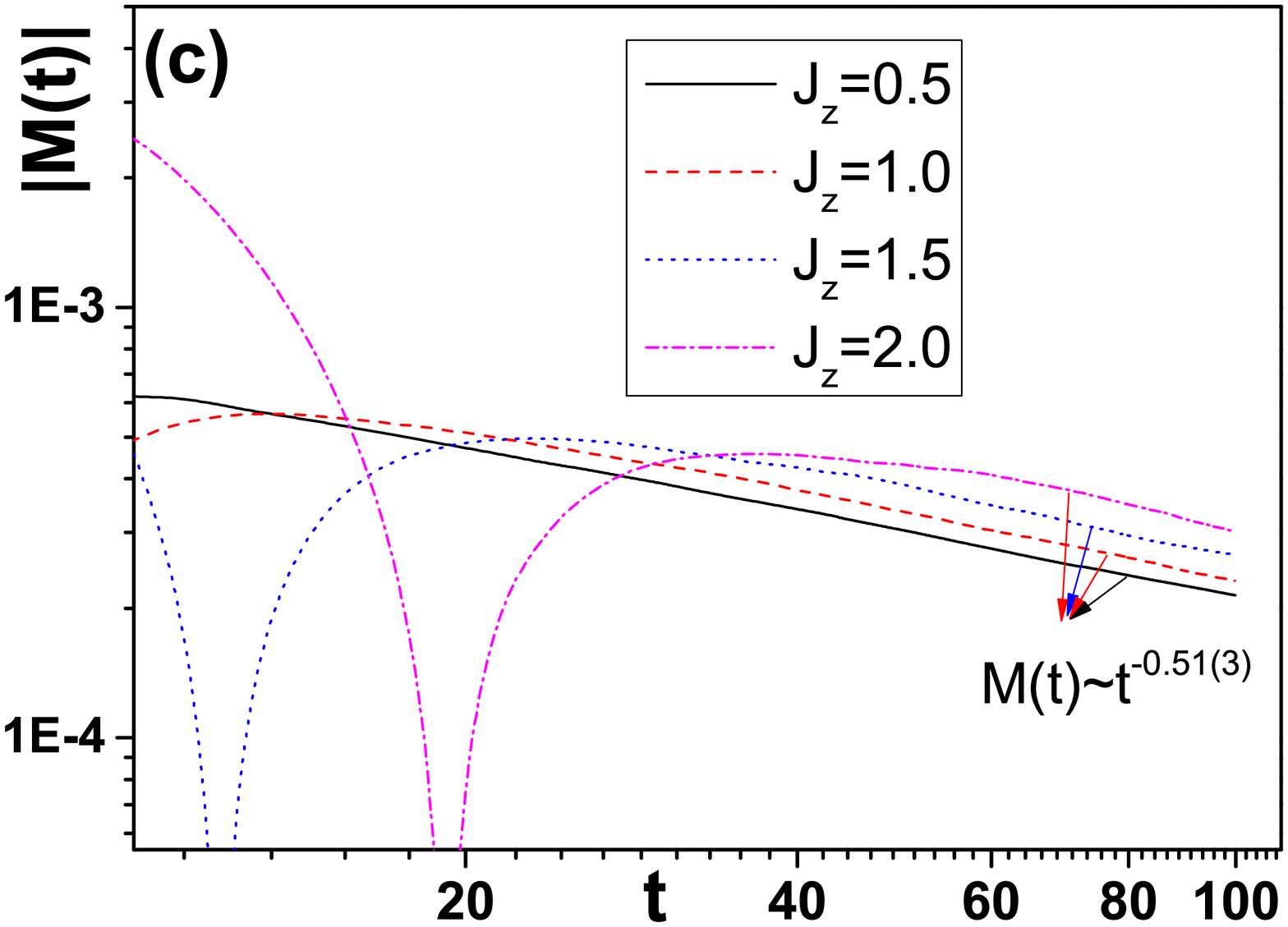}
\caption{(Color online).
Dynamics of $M(t)$  in (a) small systems with open boundary condition(OBC), the inset is the decay rate as a function of $1/L^2$; (b) large systems with a staggered external potential $h_i=V_0(-1)^i$ and (c) interacting systems with various $J_z$. $J_z=0$ for (a) and (b) and $\Delta=\delta=0$ for (a)-(c); The system size $L=320$ for (b) and $L=96$ for (c).} \label{fig:fig1}
\end{figure*}
{\it Model:} We study an 1D disordered spin-$\frac12$ model with dephasing noise: $
H(t)=H_s+\sum_i [h_i S_i^z+\xi_i(t) S_i^z]$,
where $h_i$ denotes the static disordered fields along the z-direction $h_i\in [-\Delta,\Delta]$. The noise is modeled by a stochastically fluctuating magnetic field $\xi_i(t)$ satisfying $\langle\xi_i(t)\rangle=0$ and $\langle\xi_i(t)\xi_j(t')\rangle=\gamma\delta(t-t')\delta_{ij}$. $H_s$ is an anisotropic quantum magnetic (XYZ) Hamiltonian:
\begin{small}
\begin{equation}
H_s=\sum_i (J+\delta)S_i^xS^x_{i+1}+(J-\delta)S_i^yS^y_{i+1}+J_z S_i^zS_{i+1}^z,  \label{eq:Ham}
\end{equation}
\end{small}
In the model, there are four different types of parameters in terms of $J$: the noise strength $\gamma$  fixed as $\gamma=J$ throughout the paper;  the disorder strength $\Delta$; the neareast neighboring(NN) ``interaction'' $J_z$ and the in-plane anisotropic parameter $\delta$ determining the symmetry of the model; As we will show below, despite the complexity of the model,  the interplay between the noise, disorder, interaction and symmetry could give rise to universal long-time dynamics that independent of the system details.

{\it Methods:} There are two kinds of random variables $h_i$ and $\xi_i(t)$, ensemble average needs to be perform over both variables. Average over noise trajectories $\{\xi_i(t)\}$ can be performed either explicitly\cite{Daley2014} or implicitly\cite{Gardiner1999}, the latter leads to Markovian-Lindblad master equation\cite{Lindblad1976}:
\begin{equation}
\frac{d\rho_h}{dt}=i[\rho_h,H_s+\sum_i h_i S_i^z]+\gamma \sum_i \{{S_i^z \rho_h S_i^z}-\frac 14 \rho_h\} \label{eq:Master}
\end{equation}
where $\rho_h$ is the density matrix for a given set of disorder realization $\{h_i\}$.   The ensemble average over the static disorder can be further performed  by directly sampling over 50 sets of disorder realizations $\bar\rho=\langle \rho_h\rangle_{\{h_i\}}$ and the average value of the physical quantities are defined as $\langle \hat O(t)\rangle=Tr[\hat{O}\bar\rho(t)]$. We choose the initial state as the Neel state (this specific choice does not affect our conclusion, see Supplementary Material(SM) for details~\cite{Supplementary})). Two different kinds of quantities have been caclulated: the off-diagonal correlation $C(t)=|\langle S_{\frac L2}^+S_{\frac L2+1}^-\rangle|$ and the antiferromagnetic order parameter $M(t)=\langle\frac 1L\sum_i (-1)^i S_i^z\rangle$. Even though both of them will approach zero eventually, their dynamics can be very different even for the same system parameters.

In general, it is nontrivial to study the time evolution of such an quantum many-body system. For a 1D system, however, the time-dependent density matrix renormalization group (tDMRG) algorithm\cite{Vidal2004,Luo2003,White2004,Daley2005}  has provided an unbiased numerical method that enable us to study the Master Eq.(\ref{eq:Master}) for a large system (up to a hundred system sites) and for a sufficiently long time, which is crucial for us to observe the universal dynamics that are inaccessible by the exact diagonalization(ED) method.

\begin{figure*}[htb]
\includegraphics[width=0.32\linewidth,bb=76 49 714 510]{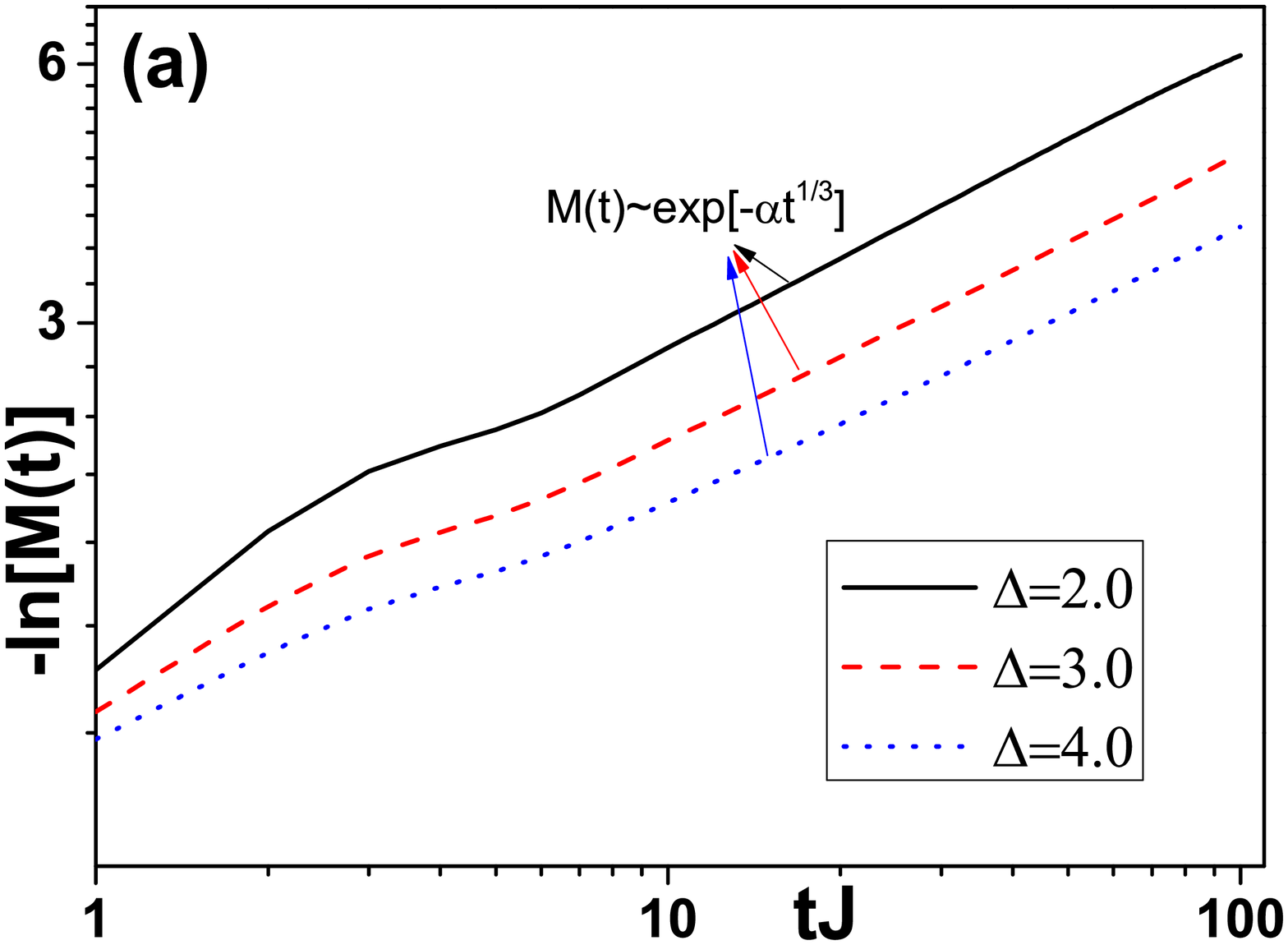}
\includegraphics[width=0.33\linewidth,bb=48 50 732 531]{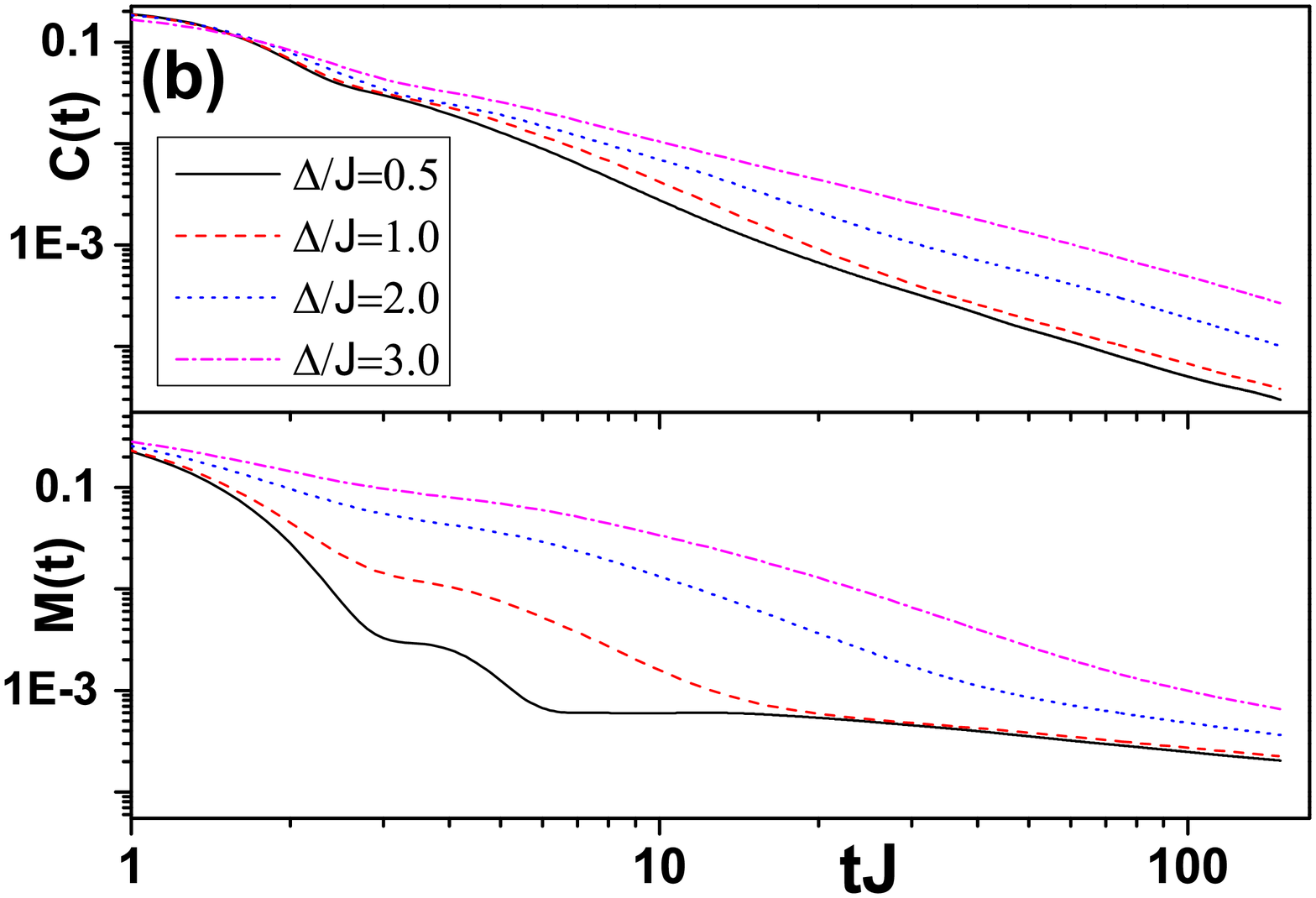}
\includegraphics[width=0.33\linewidth,bb=48 50 733 531]{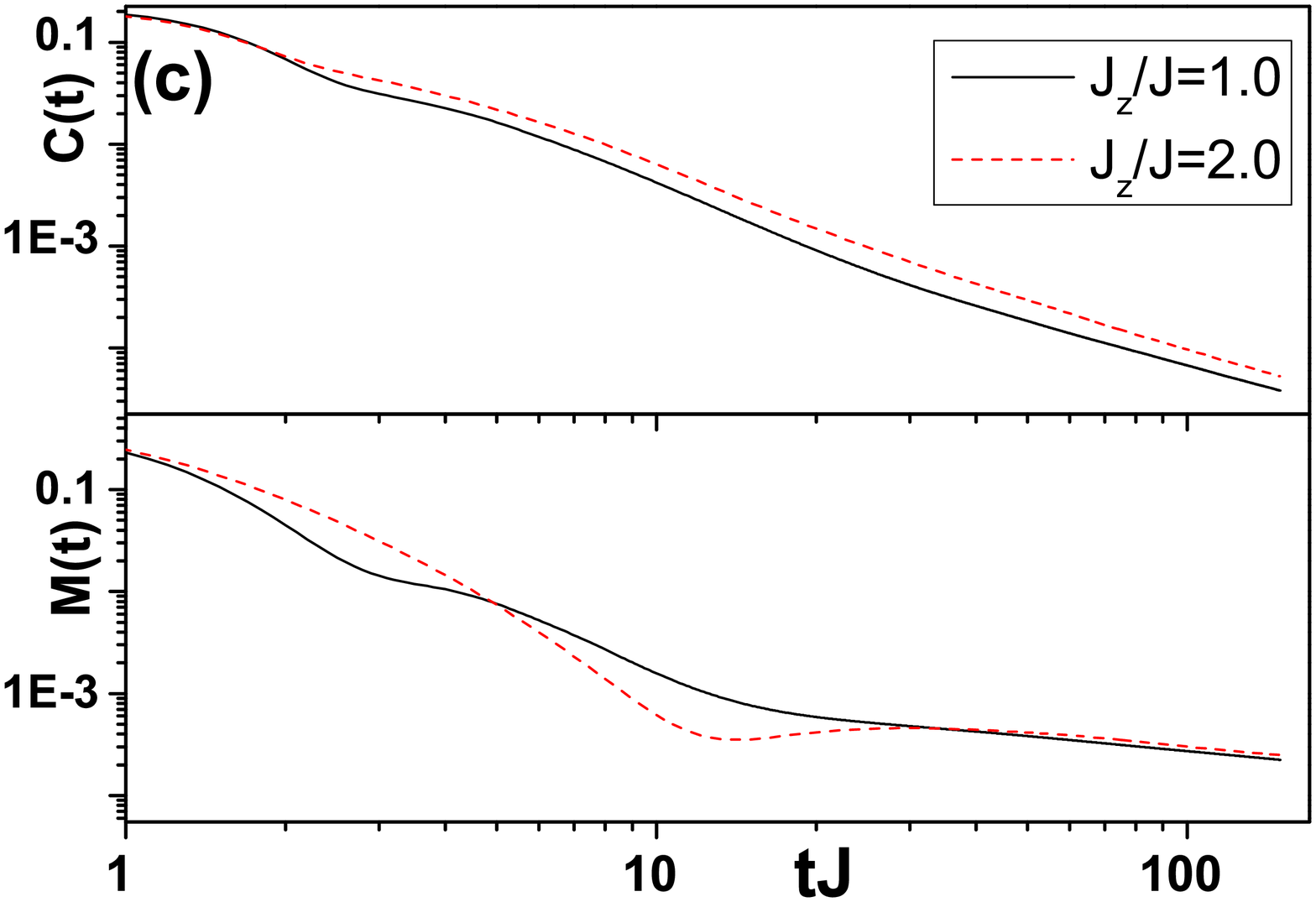}
\includegraphics[width=0.325\linewidth,bb=48 49 732 530]{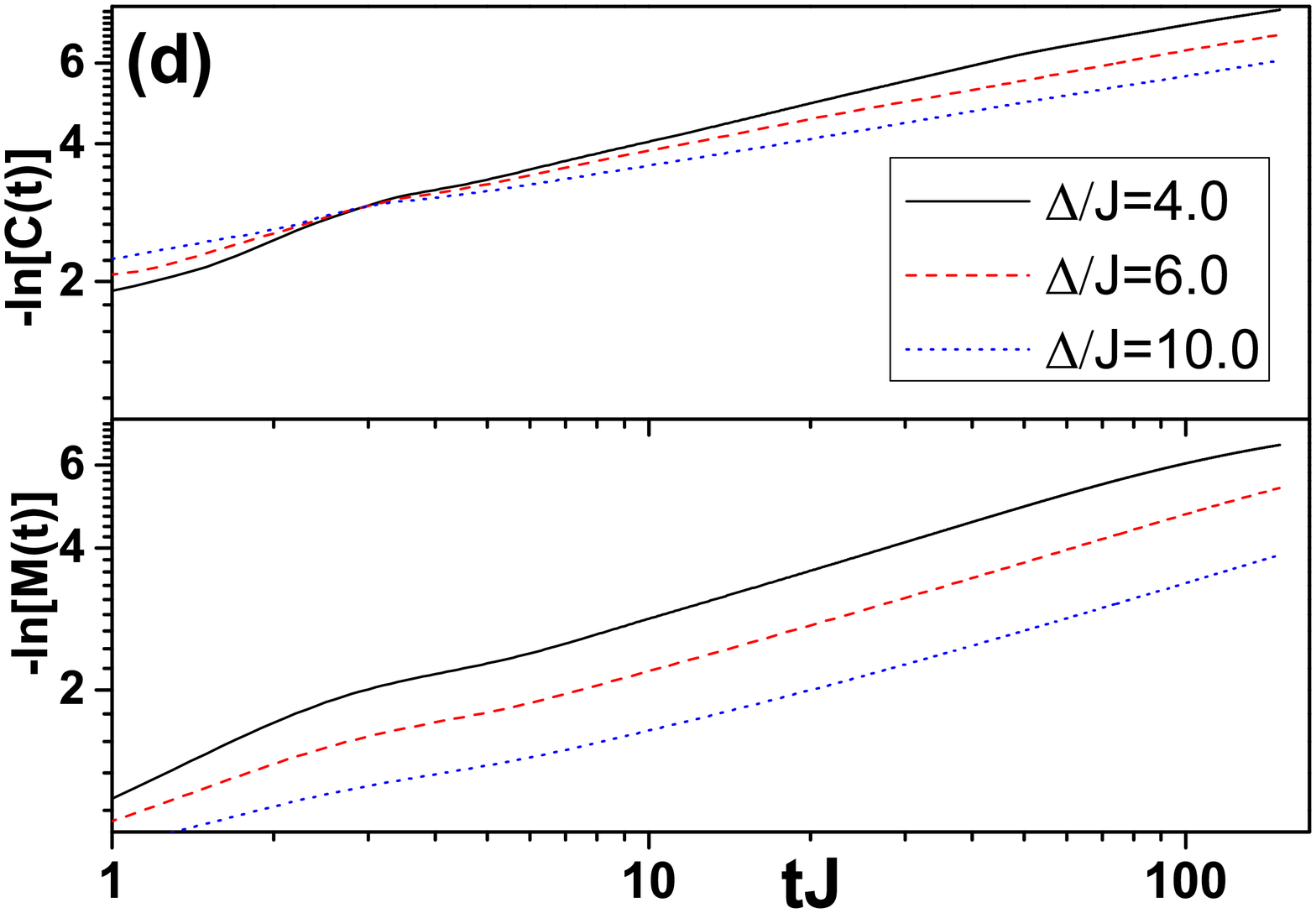}
\includegraphics[width=0.325\linewidth,bb=48 50 732 531]{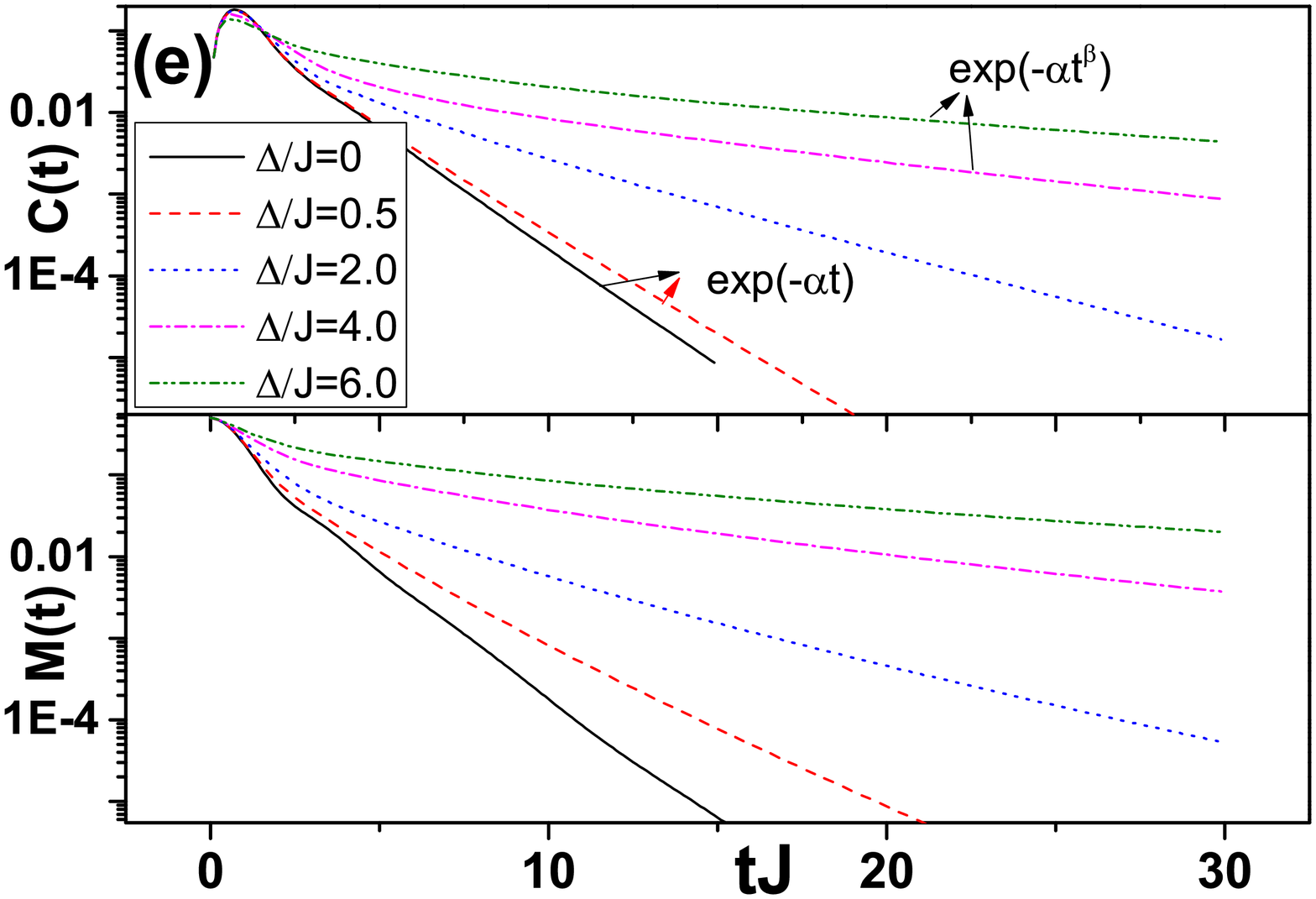}
\includegraphics[width=0.325\linewidth,bb=48 50 733 531]{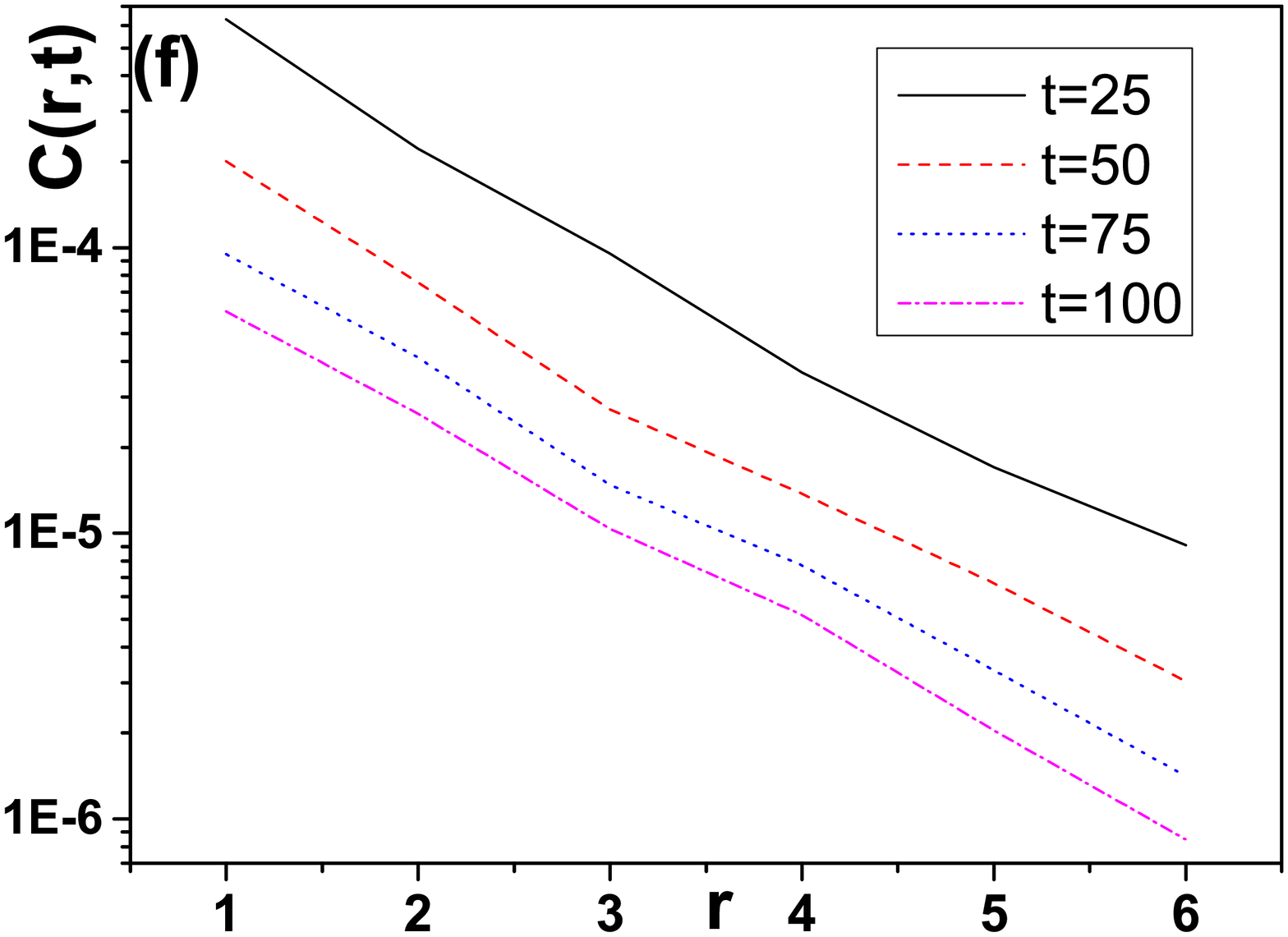}
\caption{(Color online).
(a)Dynamics of $M(t)$ with various disorder strength $\Delta$ in an noninteracting ($J_z=0$) model with lattice length $L=320$). Dynamics of $C(t)$ (upper) and $M(t)$ (lower panels) with:  various $\Delta$ but fixed interaction ($J_z=J$) in cases of (b) weak and (d) strong disorder;  (c) different $J_z$ but fixed $\Delta=J$; (e) U(1) symmetry breaking $\delta=J_z=J$. (f) The equal-time correlation function $C(r,t)$ at different time with $\Delta=J_z=J$. $L=96$ for (b)-(f), and the anisotropy $\delta=0$ except for (e).} \label{fig:fig2}
\end{figure*}

{\it Models with U(1) symmetry:} Since the conservation law is important for DUCs of classical systems\cite{Hohenberg1977}, one may wonder whether this is also the case in quantum systems. To study this point, we first focus on the case with U(1) symmetry ($\delta=0$), where the total magnetization $\sum_i S_i^z$ is conserved during the evolution. To begin with, we consider the simplest case ($\Delta=J_z=0$). Under periodical boundary condition(PBC), the system is translational-invariant thus can be decoupled into different momentum(k)-modes, each of which evolves independently with an identical decay rate $\gamma$. Therefore, both $M(t)$ and $C(t)$ decay exponentially with time ($\sim e^{-\gamma t}$). This exponential decay, however, is unstable to general mode coupling perturbations such as external potential, disorder and interaction, which will qualitatively change the long-time dynamics, as we will see below.

The simplest mode-coupling mechanism is changing the boundary condition from PBC to OBC, where boundary scattering mixes different k-modes. For a finite system with length $L$, as shown in Fig.1 (a), the initial fast decay $\sim e^{-\gamma t}$ eventually  gives way to a much slower exponential decay $\sim e^{-\Delta(L) t}$, whose decay rate decreases over the system size as $\Delta(L)\propto L^{-2}$\cite{Znidaric2015}. In a thermodynamical limit, the relaxation time diverges and algebraic decays ($M(t)\sim t^{-0.5}$, $C(t)\sim t^{-1.5}$) emerge after a sufficiently long time, as shown in Fig.\ref{fig:fig1} (b). These algebraic decays are universal in the sense that their exponents are robust against perturbations such as external potentials or interactions (even long-range interaction, see SM~\cite{Supplementary} and Ref.\cite{Landig2016}), as shown in Fig.\ref{fig:fig1}(b) and (c) respectively.

Despite the stability against the aforementioned perturbations, the algebraic decays can be qualitatively changed by relevant perturbations such as static disorder.  To demonstrate this, we first focus on the noninteracting case ($\Delta\neq 0$ and $J_z=0$). Without noises and interaction ($\gamma=J_z=0$), all the eigenstates of a 1D system are expected to be  localized\cite{Anderson1958}. External noises keep pumping energy into the system, thus finally delocalizing\cite{Gopalakrishnan2017} and driving the system towards an ITS  where all the eigenstates are equally mixed.   It turns out that the disorder may be a relevant perturbation for the asymptotic dynamics towards the ITS.   As shown in  Fig.\ref{fig:fig2}(a), one can see that the static disorder will change the long-time dynamics from an algebraic to a stretched exponential decay ($\sim \exp[-ct^{\beta}]$), where the exponent $\beta\approx 1/3$ barely depends on disorder strength. This stretched exponential decay and its exponent can be understood based on a toy model, where the onsite disorder $h_i$ can only take two values $0$ or $\infty$ with probabilities $p$ and $1-p$ respectively ($0<p<1$). In this case, a 1D chain is broken into a set of disconnected segments with OBC. The decay rate of each segment decreases over it lengths as $\Delta(l)\sim l^{-2}$. Thus, the long-time dynamics is dominated by those long segments, which, are rare events because the probability of finding a segment with length $l$  decays exponentially with $l$ ($P(l)= p^l$) in our model. Therefore, the stretched exponential decay can be understood as a collective behavior of these rare events: $O(t)\sim \int dl P(l)e^{-\Delta(l)t}\sim e^{-\alpha' t^\frac 13}$. This integral can be estimated by replacing the integrand by its value at the maximum point $l^*\sim t^{\frac 13}$.

The situation is further complicated in the presence of interactions($J_z\neq 0$, $\Delta\neq 0$). Without noises, a 1D interacting system is known to be robust against weak disorders, and it experiences an ergodic-to-MBL phase transition\cite{Pal2010}. With noises, previous studies focused on the strong disorder regime based on either the ED of small systems\cite{Levi2016,Medvedyeva2016} or quasi-classical approximation of strong decoherence case\cite{Fischer2016,Everest2017}. In the weak disorder regime, however, the long-time dynamics is sensitive to the finite size effect: the ``universal'' dynamics of small systems give way to  nonuniversal size-dependent dynamics. For sufficiently large systems, we find that the interactions recover the universal algebraic decays $M(t)\sim t^{-0.5}$, $C(t)\sim t^{-1.5}$, whose exponents depend neither on disorder (Fig.\ref{fig:fig2}b), nor on interaction(Fig.\ref{fig:fig2}c). In other words, even though a weak disorder alone is a relevant perturbation, it may become irrelevant in the presence of interaction. Nevertheless, this universal dynamics can be qualitatively changed if the disorder is strong enough to overcome the interaction effect and recovers the stretched exponential decays(Fig.\ref{fig:fig2}d), similar to the noninteracting case. The distinct relaxation dynamics in weak and strong disorder regimes indicate a dynamical phase transition, which can be characterized only by dynamical properties instead of static ones (steady states in both regimes are the same). The transition point seems to agree with that in the ergodic-to-MBL transition without noises($\Delta_c\approx 3.6J$), which can also occur at an infinite temperature. However, these two transitions are different because the coupling to noises delocalizes the system even in the strong disorder regime, thus making their dynamical properties (transport, relaxation, etc) fundamentally different from those in the noiseless case.

\begin{table*}[htb]
\begin{tabular}{|p{2cm}<{\centering}|p{2cm}<{\centering}|p{2cm}<{\centering}|p{2.3cm}<{\centering}| p{2.3cm}<{\centering}|p{2.3cm}<{\centering}|p{2.3cm}<{\centering}|}
\hline
External potential & interaction & static disorder & interaction+ weak disorder & interaction+ strong disorder & U(1) breaking+ weak disorder & U(1) breaking+ strong disorder \\ \hline
$\delta=0$, $J_z=0$ $\Delta=0$, $\gamma>0$ & $\delta=0$, $J_z>0$ $\Delta=0$, $\gamma>0$ &$\delta=0$, $J_z=0$ $\Delta>0$, $\gamma>0$ & $\delta=0$, $J_z>0$ small $\Delta$, $\gamma>0$ &$\delta=0$, $J_z>0$ large $\Delta$, $\gamma>0$ &$\delta>0$, $J_z>0$ small $\Delta$, $\gamma>0$  & $\delta>0$, $J_z>0$ large $\Delta$, $\gamma>0$  \\ \hline
 universal algebraic& universal algebraic & stretched exponential & universal algebraic & stretched exponential & non-universal exponential & stretched exponential   \\ \hline
\end{tabular}
\caption{Relaxation dynamics in our model and their dependence on disorder, interaction, symmetry and their interplay.} \label{Table:1}
\end{table*}

{\it Model without U(1) symmetry.} Up until now, we  examined the effects of various mode-coupling mechanisms. Now we address another important factor-the conservation laws, whose role can be examined by comparing the U(1) symmetric results ($\delta=0$) to those without the total-$S_z$ conservation($\delta\neq 0$). As shown in Fig.\ref{fig:fig2}(e), in the weak disorder regime, the relaxation dynamics are immediately changed from the algebraic to exponential decay as long as the total-$S_z$ conservation is broken, indicating that a system with conservation law typically relaxes much slower than that without it. For a strong disorder, we again observe a stretched exponential decay as a accumulative behavior of exponential decays with multiple decay rates. In  summary, the symmetry-breaking perturbations qualitatively alter the algebraic dynamics at a weak disorder, while it leaves the stretched exponential decay unchanged in the strong disorder regime.

{\it Discussion:} The algebraic decay of M(t) as well as the role of the conservation law can be understood in the strong noise limit, where the off-diagonal terms of $\bar{\rho}(t)$ decay much faster than the diagonal ones, thus can be adiabatically eliminated, which leads to a set of classical rate equations for the diagonal elements of $\bar{\rho}(t)$. Furthermore, with the help of conservation law, one can propose a hydrodynamics for the coarse density of spin: $\phi(x)=\sum_{i\in \mathbb{X}}S_i^z$, where $\mathbb{X}$ is a ``fluid cell'' centered at $x$. The general equation of hydrodynamics take the form:
\begin{equation}
\partial_t\phi(x,t)=-\nabla^2\phi(x,t)+\mathcal{F}(\phi(x,t)) \label{eq:diff}
\end{equation}
where $\mathcal{F}(\phi(x,t))$ represents perturbations induced by the interaction, disorder or other factors.  If they are irrelevant, Eq.(\ref{eq:diff}) describes a  diffusive process satisfying: $\phi_k(t)\propto e^{-k^2 t}$, with $\phi_k(t)=\int dx e^{ikx}\phi(x,t)$.  The dynamics of $M(t)$ can be considered as a collective behavior of different k-mode, thus $M(t)\sim \int dk \phi_k(t)\propto t^{-\frac 12}$, a signature of diffusive dynamics, which can be further verified from the divergence of its correlation length with time (see SM for details~\cite{Supplementary}).   Relevant perturbations might qualitatively change such diffusive dynamics. For instance, the effect of a perturbation breaking the conservation can be understood by adding a term  $-\gamma\phi(x,t)$ to the right side of Eq.(\ref{eq:diff}), which immediately changes the algebraic decay to an exponential one.  Generally, an explicit derivation of $\mathcal{F}$ from our microscopic Hamiltonian and a renormalization group analysis of them are beyond the scope of this work, and will be left for the future.

In spite of the diffusive nature of the dynamics of $M(t)$, the dynamics of $C(t)$ is more intriguing. On one hand,  throughout the paper we were concerned with a moderate noise regime with $\gamma=J$, where the typical amplitudes of $C(t)$  are compatible to those of $M(t)$, thus cannot be adiabatically eliminated thus the quasi-classical approximation is invalid. Such off-diagonal terms of the $\bar{\rho}(t)$ represent quantum fluctuations, whose dynamics should be treated on the same footing as the diagonal ones. For instance, the relation $\frac{d\langle S_i^z\rangle}{dt}\propto\Im {\langle (S_{i-1}^++S_{i+1}^+)S_i^-\rangle}$ indicates that $\Im\langle S_{\frac L2}^+S_{\frac{L}{2}+1}^-\rangle\propto t^{-\frac 32}$, provided that $\langle S_i^z\rangle\propto t^{-\frac 12}$ as analyzed above.  On the other hand, the quantum fluctuation dynamics itself is fundamentally different from the classical ones.  The algebraic relaxation indicates a divergence of relaxation time. However, the correlation length, which can be derived from the equal-time correlation function $C(r,t)=Tr[S_{\frac L2}^+S^-_{\frac L2+r}\bar\rho(t)]$, approaches  a constant $\xi_0$ in the algebraic dynamical regime as shown in Fig.\ref{fig:fig2}(f). $C(r,t)$ can be factorized into spatial and temporal parts as: $C(r,t)\sim t^{-\frac 32} e^{-\xi_0 r}$. The absence of correlation length divergence indicates  the dynamics of quantum fluctuations are neither diffusive nor sub-diffusive. Here, we only study the time-evolution of the equal-time correlations, while unequal-time correlations could exhibit richer dynamical behaviors that beyond the scope of this paper({\it e.g.} aging\cite{Sciolla2015,Wolff2019}), which deserves further studies.

{\it Experimental realization and detection:} Our model can be experimentally simulated by loading dipolar atoms (e.g. $^{168}$Er or$^{174}$Yb) into a deep optical lattice, where strongly repulsive interactions give rise to the hardcore nature of the bosons, thus allowing us to map it to a spin-$\frac 12$ model. The NN interactions ($J_z$ terms) have been observed in a dipolar bosonic setup\cite{Baier2016}. Both the disorder and noise can be introduced into an optical lattice via controllable ways: the former has been realized by implementing a speckle potential formed by interfering lasers\cite{Billy2008,Roati2008}, while the latter can be introduced via spontaneous light scattering\cite{Pichler2013,Luschen2017}, which can be considered as continuous local density measurements. For the detections, the density imbalance $M(t)$ can be measured by a superlattice band-mapping technique\cite{Trotzky2012} and the NN coherence $C(t)$ can be extracted from momentum distributions obtained by the time-of-flight technique\cite{Bouganne2019}.

Now we discuss the effect of various experimental imperfections on our system. First, the finite temperature effect is not a obstacle in our case, because we are interested in the physics near the ITS so an extremely low temperature is not necessary. Moreover, the algebraic dynamics is universal such that the external (harmonic) potential does not make a qualitative difference (see SM). What really matters in our case is the particle loss dissipation, which is inevitable in the cold atom setups. It clearly breaks the particle number conservation, thus qualitatively change the relaxation dynamics. However, it does not preclude the possibility of observing the algebraic decays experimentally. What matters is the time scale: as long as the particle-loss rate is much smaller than the noise strength and other system parameters, the universal algebraic dynamics can still be observed in an intermediate regime before the particle loss mechanism finally takes over (see SM). Therefore, the major challenge for the experimental realization is to control the particle loss rate, and make it much smaller than other system parameters.

Finally, we discuss the relationship between our results and two recent experiments.  In Ref.~\cite{Bouganne2019}, the dynamics of an interacting bosonic gas subjected to a near-resonant laser-induced noise has been measured and an algebraic decay of the NN correlation $C(t)\sim t^{-\beta}$ has been discovered, even though the exponent observed there ($\beta=\frac 12$) differs from that in our simulation ($\beta=\frac 32$). We expect that several factors, particularly the system dimensions, may be responsible for this discrepancy. Numerical simulations of a 2D system might be challenging, however, recent developments shed light on this direction\cite{Kshetrimayum2017}. Another possible explanation for this discrepancy is that the typical time regime studied in Ref.~\cite{Bouganne2019} is not long enough, thus the dynamics is initial-state-dependent.  In another experiment\cite{Luschen2017}, Luschen et al. observed a slow(stretched exponential) decay, which corresponds to the strong disorder regime in our case, while more interesting and universal dynamics appear in the weak disorder regime as we demonstrated. A challenging but very interesting problem for both experiments and numerics is the dynamics near the phase transition point, where the interplay between the criticality and noise may give rise to a novel relaxation behavior that differs from the dynamics on both sides of the transition point.

{\it Conclusion and outlook:} We studied the long-time dynamics of noisy quantum many-body systems, where the interplay between the interaction, disorder, noise and symmetry leads to interesting universal dynamics, as summarized in Tab.\ref{Table:1}. Several important problems are left for studies in the future, including the dynamics near the critical point and in higher dimensions, both of which call for new algorithms. We also expect similar universal dynamics can be observed for interacting fermionic systems. In addition, introducing dissipation in our model might give rise to driven-dissipative systems with interesting non-equilibrium steady states and  dynamics. Finally, the universality of the relaxation dynamics implies a unified underlying mathematical structure behind them,  which inspires us to develop an effective non-equilibrium field theory to categorize different relaxation dynamics  and determine the relevance of perturbations via controllable renormalization group analysis.

{\it Acknowledgment.}   This work is supported in part by the National Program on Key Research Project  under  No. 2016YFA0302001and 2016YFA0300501,  and the National Natural Science Foundation of China under No. 11674221, 11574200, 11974244 and 11745006. This work is also supported by the Project of Thousand Youth Talents, the Program Professor of Special Appointment (Eastern Scholar) at Shanghai Institutions of Higher Learning, the Shanghai Rising-Star program and the Shanghai talent program.


\end{document}